# Large Area Single Crystal (0001) Oriented MoS$_2$ Thin Films

AUTHOR NAMES

*Masihhur R. Laskar[†], Lu Ma[‡], ShanthaKumar K[§], Pil Sung Park[†], Sriram Krishnamoorthy[†], Digbijoy N. Nath[†], Wu Lu[†], Yiying Wu[‡], Siddharth Rajan[†*]*

AUTHORS ADDRESS

[†]Department of Electrical and Computer Engineering,

[‡]Department of Chemistry and Biochemistry,

The Ohio State University, Columbus 43210, Ohio, USA and

[§]Department of Nanobio Materials and Engineering,

Gwangju Institute of Science and Technology, Gwangju 500-712, Republic of Korea.

*CORRESPONDING AUTHOR:

E-mail: rajan@ece.osu.edu

Phone: +1(614) 247 7922.

Fax: +1 (614) 292 7596

## ABSTRACT

Layered metal dichalcogenide materials are a family of semiconductors with a wide range of energy band gaps and properties, and the potential to create exciting new physics and technology applications. However, obtaining high crystal quality thin films over a large area remains a challenge. Here we show that chemical vapor deposition (CVD) can be used to achieve large area electronic grade single crystal Molybdenum Disulfide ($MoS_2$) thin films with the highest mobility reported in CVD grown films so far. Growth temperature and choice of substrate were found to critically impact the quality of film grown, and high temperature growth on (0001) orientated sapphire yielded highly oriented single crystal $MoS_2$ films for the first time. Films grown under optimal conditions were found to be of high structural quality from high-resolution X-ray diffraction, transmission electron microscopy, and Raman measurements, approaching the quality of reference geological $MoS_2$. Photoluminescence and electrical measurements confirmed the growth of optically active $MoS_2$ with a low background carrier concentration, and high mobility. The CVD method reported here for the growth of high quality $MoS_2$ thin films paves the way towards growth of a variety of layered 2D chalcogenide semiconductors and their heterostructures.

## INTRODUCTION

In the past decade, there has been a resurgence of interest in two-dimensional and layered materials stimulated largely by the work on the unique bandstructure and properties of graphene [1-5], whose applications are still being widely investigated. A second, more diverse group of layered semiconductors is the metal dichalcogenide family, which has received significant attention recently for next generation electronics [6-11], optoelectronics [12], and sensors [8,13]. In contrast to graphene, these materials have large band gaps and can therefore be used for many electronic and optoelectronic device applications. While their mobility and velocity are comparable to Si, the intrinsic 2-dimensional nature of carriers in these materials offers the advantages of superior vertical scaling for a transistor topology [13]. The layered nature of these materials offers the exciting prospect of creating heterostructures that are not limited by out of plane bonding and lattice mismatch through techniques such as van der Waals epitaxy. From a technological viewpoint, the potential of these materials for low-cost flexible or transparent electronics could revolutionize technology.

Promising $MoS_2$ device (transistor) results showing excellent on-off ratio and high current density have been obtained using flakes [6-7, 9-11] mechanically exfoliated from bulk geological samples. While exfoliation of $MoS_2$ is valuable for demonstrating the promise of $MoS_2$ devices, it is not suitable for device synthesis where repeatable and uniform large area deposition is desirable. While previous approaches involving chemical vapor deposition [17-19], sulfurization of molybdenum oxides [20-21], hydrothermal synthesis [22] and electrochemical synthesis [23] led to large area thin $MoS_2$ the crystal quality of the layers grown was significantly lower than that obtained from naturally occurring $MoS_2$, which remains the highest quality material available to date. The low structural quality and polycrystalline nature, evident from Raman spectroscopy, X-ray diffraction, and transmission electron microscopy in these reports lead to low mobility which was orders of magnitude lower than the mobility in relatively high-quality exfoliated samples. Methods to achieve large area metal dichalcogenide films that use the conversion of bulk $MoS_2$ into large scale $MoS_2$ film using laser trimming [15] and liquid exfoliation [16] exploit the high quality of geological $MoS_2$ but have not demonstrated area-specific and large area thin film synthesis.

In this paper, we show for the first time that chemical vapor deposition on epitaxial substrates can lead to large area oriented crystalline films with unprecedented high quality. The films reported here have structural quality similar to the best quality geological samples available to date, and could enable a variety of large area electronic and optoelectronic device applications. The methods proposed can also be extended to other layered chalcogenide materials, as well as lateral and vertical heterostructures based on them.

**RESULTS AND DISCUSSION**

Samples were grown by sulfurization of e-beam evaporated Mo films (see Methods for more details). Our experiments indicate that growth temperature and substrate were critical in determining film quality. We describe here characteristics of samples grown at temperatures 500°C, 700°C, 900°C, and 1100°C, referred to as Sample A, B, C, and D, respectively. From optical microscopy, each of these films was found to have specular reflection indicating smooth $MoS_2$ coverage and absence of any remaining metal on the surface. Atomic force micrographs of the films (Fig.2) show the evolution of surface morphology as a function of growth temperature. At relatively low temperatures Sample A showed smooth surface morphology (rms roughness = 0.30) with small grains that we attribute to the polycrystalline nature of the film. For Sample B, the grain size was found to be larger, while the surface roughness (3.5 nm) was higher than Sample A. In Sample C, thin sheet of $MoS_2$ was found to cover the entire surface, and the morphology revealed hexagonal symmetry. We believe this originates from the hexagonal basal plane symmetry of the $MoS_2$ crystal itself, suggesting that the layers grown are oriented along (0001) or c-axis of the $MoS_2$ unit cell. In sample D, $MoS_2$ flakes several microns in lateral size and tens of nanometer thick were formed, indicating the significant mass transport takes place during the growth. MoS$_2$ flakes with sharp edges and corners were found on the surface with sides parallel to step-edge of sapphire substrate which is an indication that the layers perhaps are grown epitaxially. Further detailed analysis of the AFM image of Sample C shown in Fig. 3 reveals that the step edges are 0.56nm thick, approximately equal to one atomic monolayer of MoS$_2$ [9]. Spiral growth features similar to those seen in other material systems [24] were also evident. AFM measurements were also done at the boundary of the MoS$_2$ films (more details in supplementary information). Line scans from these regions (marked by red line) reveal that the thickness of the film is approximately 6.3nm, corresponding to roughly 10 monolayers.

Samples were characterized using Raman spectroscopy (Renishaw, 514 nm laser and 60mW power) to assess the presence and quality of the MoS$_2$. Raman peaks for the $E^1_{2g}$ and $A_{1g}$ vibration modes 382 cm$^{-1}$ and 407 cm$^{-1}$ were seen in both samples, and the separation indicates that the films have that the MoS$_2$ (bulk nature of the) films were bulk-like [25]. In Fig 4(b) the Raman peaks observed from Sample A and Sample B taken under identical conditions are shown. In the case of the lowest growth temperature (500°C) sample A, the intensity of $E^1_{2g}$ peak was lower than the $A_{1g}$ peak, whereas the opposite is seen in bulk geological MoS$_2$. This is similar to the ratio of the Raman peaks from previous reports on CVD [17-19], and indicative of low structural quality in the film. In Sample B, which was grown at a higher temperature, however, the overall intensity of the peaks increased and the relative ratio, of the $E^1_{2g}$ the $A_{1g}$ peaks were similar to high-quality exfoliated MoS$_2$.. Samples C and D, which were grown at even higher temperatures, showed similar ratios (i.e. $E^1_{2g}$ peak greater than $A_{1g}$), but with intensity that was 50 times higher than Sample B. This suggests that structural quality in these films was significantly higher than Samples A or B and that higher growth temperatures are critical for the formation of high quality MoS$_2$. In Fig.4b, Raman measurements of Sample C and thick flakes from a commercial SPI $MoS_2$ wafer are compared. With equal laser power (1% laser power), the width and intensity of the peaks are almost identical. The full width at half maximum (FWHM) of the Raman peak can be correlated with the quality of the film, and as expected (Figure 4c) the FWHM decreases with increasing temperature of sulfurization, approaching that of commercial $MoS_2$. The intensity ratio of $E^1_{2g}$ and $A_{1g}$ peaks as a function of growth temperature for Samples A-D, and for a commercial bulk geological sample is shown in Figure 4(c). Samples C and D exhibit a ratio >1, similar to bulk $MoS_2$ indicating similar structural quality from Raman spectroscopy.

X-ray diffraction measurements of the $MoS_2$ confirm the trends evident from the Raman spectra. High resolution ω-2θ scans of the four investigated Samples A, B, C, D, and for commercial (geological) bulk $MoS_2$ are compared in Fig 5a. In the reference bulk sample, a sharp (0002) diffraction peak was observed at 2θ = 14.5, corresponding to the (0002) diffraction condition, Higher order peaks for the $MoS_2$ and the (0006) diffraction peak of the sapphire substrate (blue) were observed. In the case of the CVD grown thin layers, while Sample A showed no clear diffraction peaks, the lowest order peak (0002) peak with 2θ = 14.5) was clearly visible in the spectra for Sample B. In the case of Samples C and D, higher order peaks are evident from the figure, indicating long-range crystalline order in these samples. It is evident from these results that increasing temperature improves structural quality and crystal orientation of the $MoS_2$. In triple axis X-ray rocking curve measurements about the (0002) reflection condition for sample C was done to assess the sample quality and the FWHM was found to be 25 arc sec, confirming that the films are highly oriented with negligible disorder along the stacking axis. High resolution ω/2θ scans (Figure 5c) show distinct X-ray thickness fringes due to the sharp interface between $MoS_2$ film and sapphire substrate. The films grown were too thin to observe any off- axis peaks that would have helped to determine the epitaxial relation with the sapphire substrate. The X-ray diffraction measurements confirm the excellent quality of these films.

HR-TEM measurements revealed ordered crystalline $MoS_2$, confirming the XRD measurements. Selected area electron diffraction (SAED, Figure 6b) pattern taken from the MoS2/sapphire interface indicate that the CVD grown $MoS_2$ is hexagonal and oriented along the C (0001) direction of sapphire. The epitaxial relationship between the basal planes of $MoS_2$ and sapphire could not be determined from this measurement. However, though there were few slightly rotated regions, no amorphous region were observed confirming the crystalline nature of the $MoS_2$ over a large area. Figure 6a, shows the $MoS_2$ and $Al_2O_3$ growth direction along [0001] direction and the ordering of the $MoS_2$ layers is uniform periodic atom arrangement over a large area.

To investigate the effect of substrate on the $MoS_2$ growth, films were grown on $SiO_2$/Si substrates with growth conditions identical to Sample C, which as discussed earlier in this paper, had excellent structural quality. The XRD spectra of the films grown on 90 nm $SiO_2$/p-Si substrates did not show any peaks, indicating that they were either polycrystalline or amorphous, matching reports for lower temperature CVD growth of $MoS_2$ on $SiO_2$. We conclude that the hexagonal symmetry of the basal plane in sapphire plays an important role in determining the crystalline quality and orientation of the overgrown $MoS_2$ layers. Other crystalline substrates such as GaN and ZnO, or other 2D materials, could also be viable candidates for high quality $MoS_2$.

The transport properties of these films were investigated by depositing Ti/Au contacts using photolithography, metal evaporation and lift-off. Current voltage (I-V) characteristics of Sample C for different contact spacings (L) are shown in Fig. 7. The I-V characteristics, shown in a log-log scale display two different regimes with dependence I α V at lower current density, and I α $V^2$ at higher current density. The transition of the voltage dependence of current from linear to quadratic is a characteristic feature of space charge limited transport [26], suggesting that the $MoS_2$ has relatively low background carrier density, a desirable property for field effect transistors, and that current is carried through injection of electrons from the contacts into a nominally insulating semiconductor.

While the well-known Mott Guirney law for space charge transport in semiconductors, $I = 2\varepsilon_0\varepsilon_r\mu V^2/\pi L^2$ is applicable to bulk transport where 1-dimensional electrostatics are valid, it is not applicable in the present case where the field is applied laterally. We therefore follow previous work [27-28] on the analysis of space charge transport in thin films with a lateral contact geometry. The current density (in A/cm) in this case is given by $I = 2\varepsilon_0\varepsilon_r\mu V^2/\pi L^2$, where the current still has quadratic dependence on the voltage, as in the Mott-Guirney equation, but the dependence on the distance between the contacts is now modified from I α $V^2/L^3$ to I α $V^2/L^2$ due to the thin film geometry. Our analysis shows that indeed, the dependence of the current density in the space-charge region varies as $V^2/L^2$ (thin film transport) rather than $V^2/L^3$ (bulk transport). Based on this equation, we estimate field effect mobility independently from each of the curves, a high field-effect mobility 12 +/- 2 $cm^2$/V.s in the $MoS_2$ films. Gated measurements showed that the current increased as the gate voltage was made more positive, suggesting that the conduction is n-type.. Based on previous bandstructure calculations [29], the effective mass of electrons and holes in $MoS_2$ are not expected to be significantly different, and therefore, electron and hole mobility may be expected to be similar in magnitude. Using the mobility extracted from the quadratic region and conductivity from the linear region, and a carrier density $10^{10}$ cm$^{-3}$ was estimated.

The mobility reported here is higher than previous reports of CVD grown MoS$_2$, but the estimate from space charge injection is likely to be much lower than the low-field mobility that would be applicable if degenerate gases were induced in these films. Firstly, space charge injection takes place at relatively high fields where the mobility is usually lower in most semiconductors due to increased energy dissipation mechanisms such as optical phonon emission. Secondly, due to the low background charge, screening, which plays a very important role in mitigating charged defect related scattering [30] in doped and degenerate carrier gases, is expected to be minimal. Since the film is thin, we expect that remote charge scattering can degrade mobility significantly especially when there is little or no screening of the impurity scattering potential. In the future, degenerate 2D gases or higher density gases through doping in these films could be expected to have higher mobility due to the screening effects.

**CONCLUSION**

Large area (0001) oriented crystalline MoS$_2$ films with structural quality comparable to bulk geological samples were achieved using sulfurization of Mo. We find that the use of crystalline substrates and suitable growth conditions allows for synthesis of high quality crystalline films with mobility significantly higher than previous results. The method could be extended to synthesize other members of the layered metal dichalcogenide family, such as TiS$_2$, WS$_2$, and HfS$_2$, many of which are promising for electronic and optoelectronic applications. Depositing constituent metals on top of each other, or adjacent to each other could enable vertical and lateral heterostructures, and inclusion of other elements could be achieved by changing the composition of the metal film or the sulfur precursor, enabling alloys and doping.

Epitaxial growth of 2D layered materials on technologically important hexagonal symmetry semiconductors such as GaN, ZnO, and SiC could also expand the functionality of these semiconductors. The growth techniques described here fill a critical gap in taking these materials from the laboratory to real applications that need large-area high quality crystals.

**ACKNOWLEDGEMENTS**

E.L., W.L., and S.R. acknowledge funding from the NSF NSEC (CANPD) program (EEC0914790); S.R. acknowledges support from NSF ECCS Grant ECCS-0925529. L.W. acknowledges support from NSF Grants (ECCS0824170 and CMMI0928888). L. M. and Y. W. acknowledge the support from NSF (CAREER, DMR-0955471). M.L. was supported by the OSU NSF MRSEC CEM Seed Program. TEM work was supported by the World Class University (WCU) program at GIST through a grant provided by the Ministry of Education, Science and Technology (MEST) of Korea. M.L would like to acknowledge Craig Polchinski and Sanyam Bajaj for AFM measurements, and Prof. Roberto Myers and Thomas Kent for use of the photoluminescence setup.

Table 1: Calculation of mobility and carrier concentration from TLM IV characteristics $I = BV/L + CV^2/L^2$ where $B = qn_0\mu$ and $C = 2\varepsilon_0\varepsilon_r\mu/\pi$. Here $q$ = charge of an electron, $n_0$ = carrier concentration, $\mu$ = mobility, $\varepsilon_0\varepsilon_r$ = permittivity of MoS$_2$.

| TLM spacing L (μm) | Ohmic constant (B) | Space charge const. (C) | Carrier concentration n (cm$^{-3}$) | Mobility μ (cm$^2$/V.s) |
|---|---|---|---|---|
| 2 | 1.00×10$^{-5}$ | 1.03×10$^{-6}$ | 2.6×10$^{10}$ | 12.2 |
| 3 | 5.69×10$^{-6}$ | 4.37×10$^{-7}$ | 1.5×10$^{10}$ | 11.6 |
| 4 | 1.90×10$^{-7}$ | 2.60×10$^{-7}$ | 1.6×10$^{10}$ | 12.3 |

**SUPPLEMENTARY INFORMATION**

**Synthesis of MoS$_2$**: A schematic of our synthesis of MoS2 has been illustrated in Fig1. First, sapphire substrates were cleaned with solvents (acetone, methanol, 2-propanol) using ultrasonic, then baked at 120°C for 5 minutes. 50Å molybdenum metal was deposited on substrates using e-beam evaporation at a rate of 0.1 Å/s. 20 mg of sulfur was placed in a quartz-boat and the boat was placed inside a quartz tube (1cm inner diameter) along with the sample. The open end of the tube was pumped down using a mechanical pump, sealed, and placed inside a furnace. At high temperatures solid sulfur vaporizes within the tube, reacts with the metal on the sample. Several parameters can be controlled in this process including the temperature of sulfurization, time, amount of sulfur, and thickness of Mo deposited on the surface. During growth, the pressure inside the tube is determined by temperature and amount of sulfur. In our experiments, samples which were sulfurized at temperatures 500°C, 700°C, 900°C and 1100°C for 12 hours, respectively. More recent results indicate that a shorter growth time 30 minutes gives identical results. The thickness and area of MoS$_2$ film grown is limited by the thickness of the evaporated metal.

**Characterization**: Atomic force microscopy (AFM) was employed to investigate the surface morphology of the MoS$_2$ films. A Bruker AFM system with carbon tips operating at 300kHz resonant frequency was used for all scans. Raman characterizations were done using a Renishaw spectrometer with a CCD camera and laser excitation of 514 nm wavelength and 60 mW power. High resolution X-ray diffraction was done using a Bruker X-ray diffractometer with Cu Kα source. A Technai F20[F20] Philips instrument operated at 200 keV was used for TEM imaging. Cross section samples for study were prepared using a focused ion beam nano manipulator (Quanta 3D FEG, FEI). Room temperature photoluminescence was performed using a CCD camera, monochromator and a laser excitation of 300 nm wavelength and 30 mW power generated from assembly of Ti-sapphire laser and $3^{rd}$ harmonics generator. To measure electrical properties, transmission-line-measurement (TLM) structures were fabricated by optical lithography using GCA6100C stepper followed by e-beam evaporated Ti/Au (20 nm/100 nm) to form an ohmic contact onto $MoS_2$ film. The contact dimension is 30 μm × 100 μm, and the spacings between adjacent contacts vary from 2 μm to 10 μm in steps of 1μm and from 10 μm to 18 μm in steps of 2 μm. The device isolation, was done using $BCl_3/O_2$-based plasma etching using a PlasmaTherm ICP-RIE. Pulsed current-voltage measurements were done using a B1500 spectrum analyzer and an on-wafer probe station.

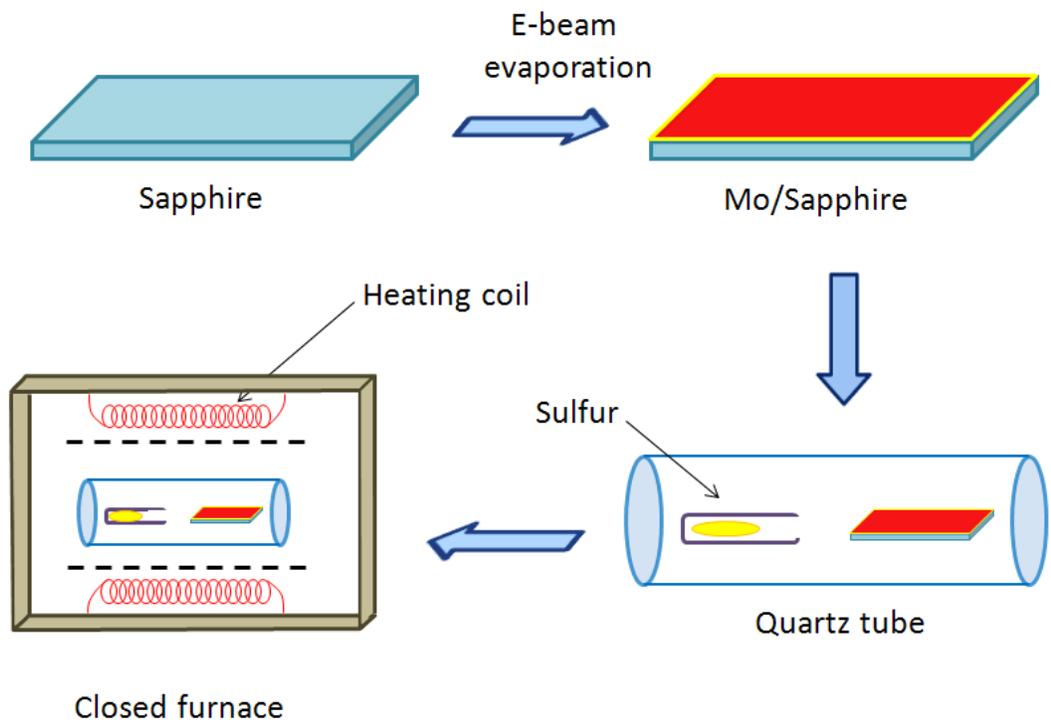

Figure 1: Schematic of the growth process.

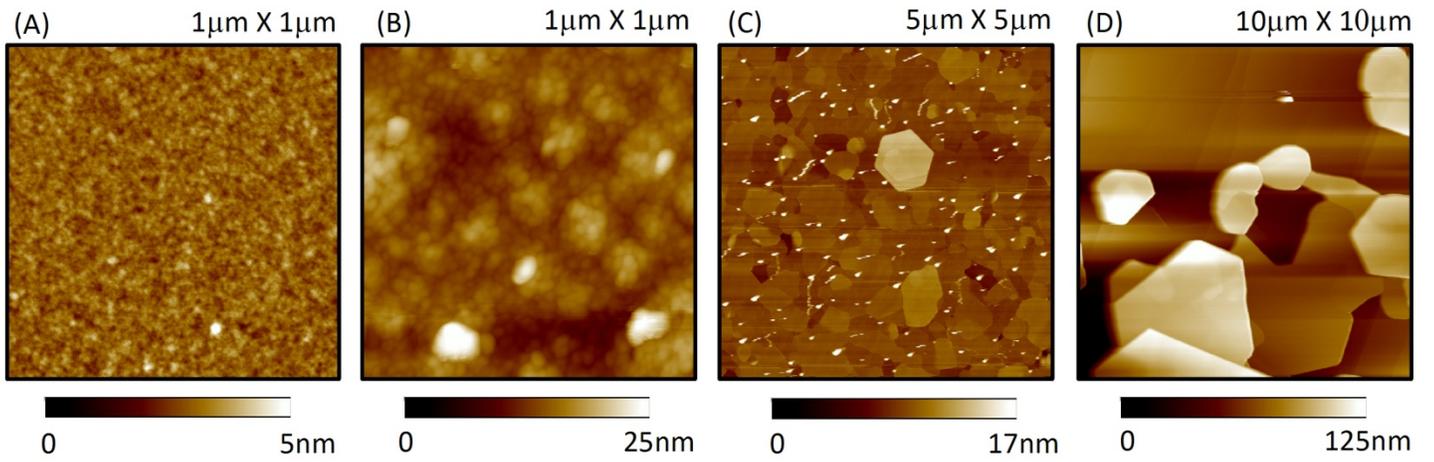

Figure 2: Atomic force micrographs of (a) Sample A (500°C), (b) Sample B (700°C), (c) Sample C (900°C), (d) Sample D (1100°C) samples. Scan area and height data scales are varied to display surface features clearly.

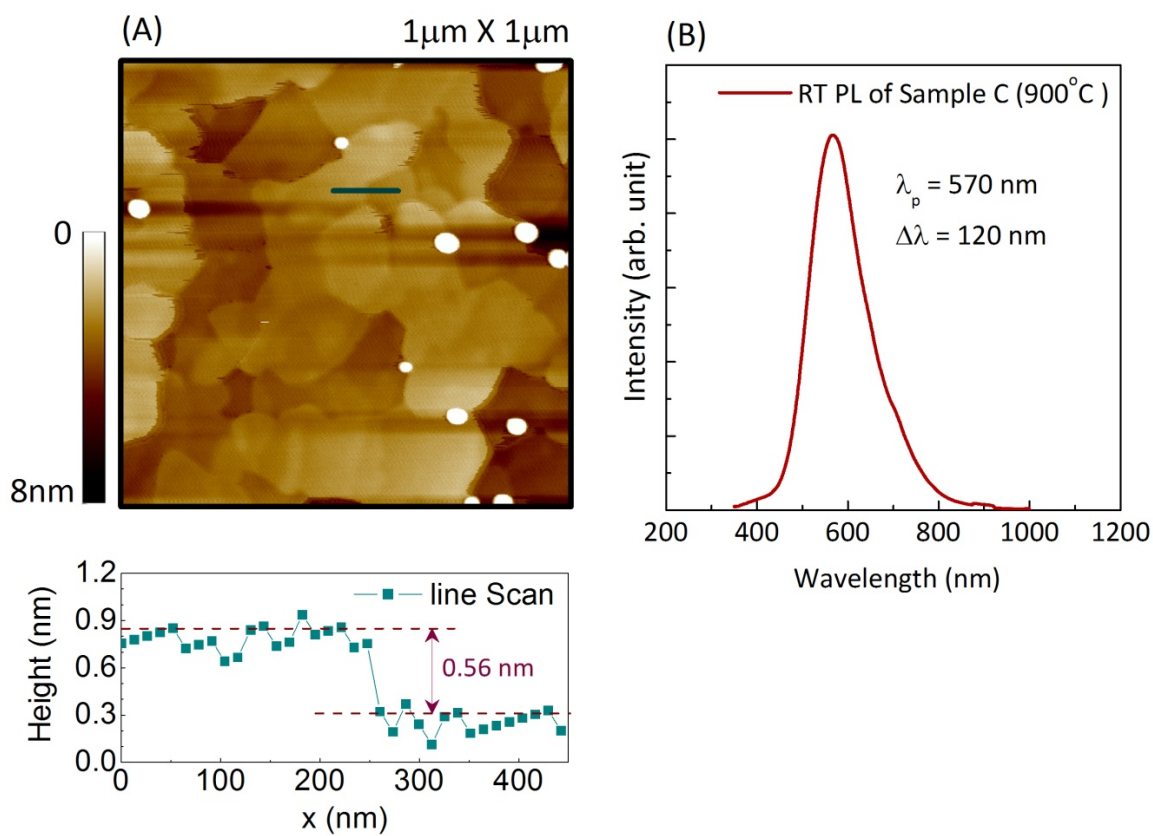

Figure 3: (A) (top) Atomic force micrograph of Sample C (900°C). A line scan (indicated by the red line) shows monolayer atomic steps on the surface. The white dots in the image are due to excess sulfur. (B) Photoluminescence of the 900°C sample shows the room temperature luminescence at 570nm.

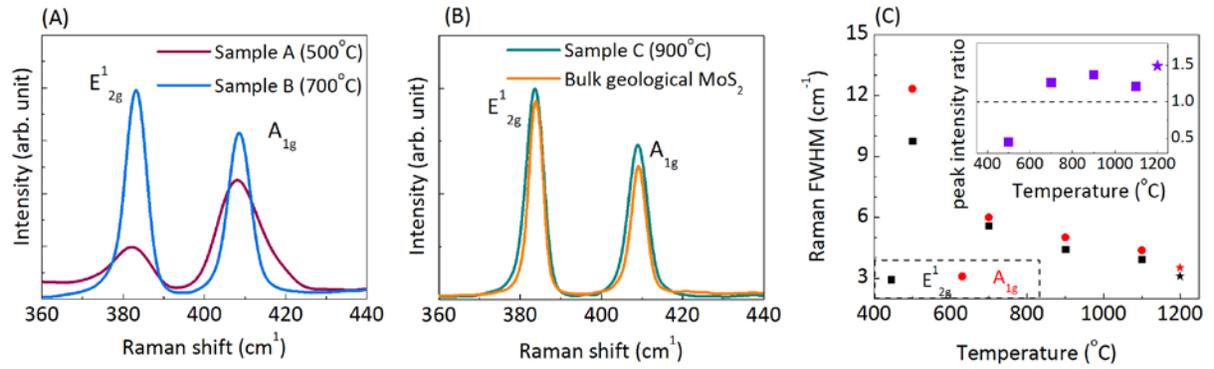

Figure 4: (left) Raman spectra of (A) Sample A (500°C) and Sample B (700°C) taken under identical conditions. The peak intensity ratio $E^1_{2g}/A_{1g}$ is less than 1 for Sample A and greater than 1 for Sample B, confirming higher growth temperature leads to improved film quality. (B) Raman spectra obtained from Sample C (900°C) and a thick (>100 nm) flake from a bulk geological $MoS_2$ show similar peak intensity and peak sharpness. Note that the peak intensity for Sample C is approximately 50 times higher than for Samples A or B. (C) Raman peak FWHM as a function of growth temperature. Increasing temperature leads to lower FWHM associated with improved structural quality. Inset: Peak ratio for $E^1_{2g}$ and $A_{1g}$ peaks for samples A, B, C, and D. The corresponding values for bulk $MoS_2$ is indicated by a star symbol.

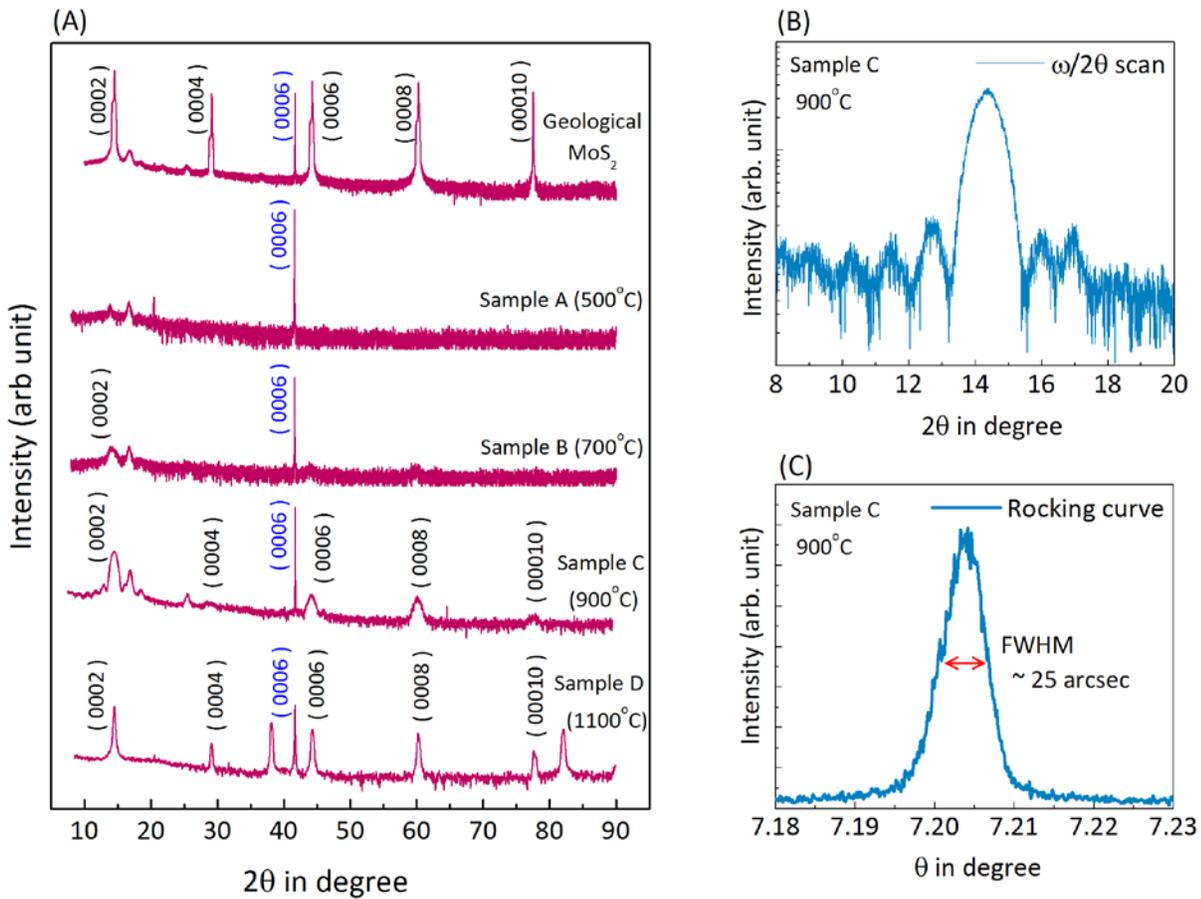

Figure 5: (a) Wide angular range X-ray diffraction spectra for Samples A, B, C, and D and bulk MoS$_2$. The diffraction peak for (0002) reflection condition for MoS$_2$ can be seen distinctly for Sample C (900°C) and Sample D (1100°C), suggesting the films are c-plane oriented (b) Triple axis ω/2θ–scan for 900°C sample shows thickness fringes implies sharp interface between MoS$_2$ film and Sapphire substrate. (C) (0002) rocking curve for Sample C indicates a low full width at half maximum of 25 arc sec.

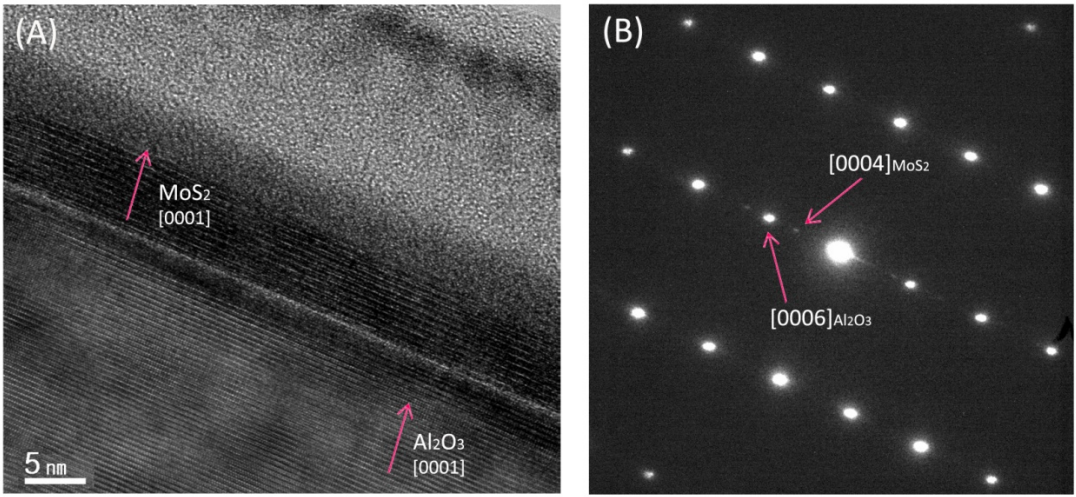

Figure 6: (A) Cross-sectional TEM image of sample C (900°C) shows perfect stacking of MoS$_2$ layers with (0001) orientation and, (B) plan view diffraction pattern from MoS2 confirms single crystalline quality of the film.

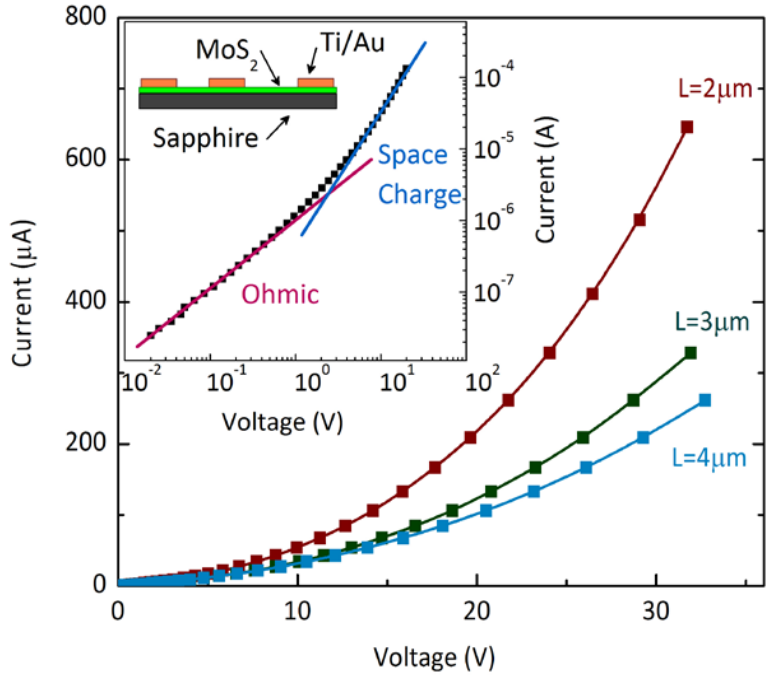

Figure 7: Current-Voltage measurements of the 100µm×25µm Ti/Au contact pads with adjacent separation of L = 2µm, 3µm and 4µm. In log-log scale two different regions can be identified based on slopes of this plots - Ohmic and space charge limited transport. Inset shows a vertical schematic and side-view image of the TLM pads.